\documentclass[12pt]{article}
\usepackage{graphicx}
\usepackage{epstopdf}
\usepackage{wrapfig}
\begin{document}
\title{From quark and nucleon correlations to discrete symmetry and clustering in nuclei}
\author{G. Musulmanbekov \\
JINR,\\
Dubna, RU-141980, Russia\\
E-mail: genis@jinr.ru
}
\date{}
\maketitle
\begin{abstract}
Starting with a quark model of nucleon structure in which the valence quarks are strongly correlated within a nucleon, the light  nuclei are constructed by assuming similar correlations of the quarks of neighboring nucleons. Applying the model to larger collections of nucleons reveals the emergence of the face-centered cubic (FCC) symmetry at the nuclear level. Nuclei with closed shells possess octahedral symmetry. Binding of nucleons are provided by quark loops formed by three and four nucleon correlations. Quark loops are responsible for formation of exotic (borromean) nuclei, as well. The model unifies independent particle (shell) model, liquid-drop and cluster models.
\end{abstract}


\section{Introduction}\label{aba:sec1}
Historically there are three  well known conventional nuclear models based on different assumption about the phase state of the nucleus: the liquid-drop, shell (independent particle), and cluster models. The liquid-drop model requires a dense liquid nuclear interior (short mean-free-path, local nucleon interactions and space-occupying nucleons) in order to predict nuclear binding energies, radii, collective oscillations, etc. In contrast, in the shell model each point nucleon moves in mean-field potential created by other nucleons; the model predicts the existence of nucleon orbitals and shell-like orbital-filling. The cluster models require the assumption of strong local-clustering of particularly the 4-nucleon alpha-particle within a liquid or gaseous nuclear interior in order to make predictions about the ground and excited states of cluster configurations. The dilemma of nuclear structure theory is that these mutually exclusive models work surprisingly well for qualitative and quantitative explanation of certain limited data sets, but each model is utterly inappropriate for application to other data sets. There is a wide variety of attempts to solve the problem of nuclear structure by the conception of binary nucleon--nucleon interactions. Even three-body forces, introduced to improve the situation, do not provide a solution.

Particle physicists believe that the fundamental theory of the strong interactions is Quantum Chromodynamics (QCD). However, the description of the dynamical structure of hadrons and, especially, nuclei in the framework of QCD has thus far remains an unsolved problem. Quark degrees
of freedom manifest themselves at high-momentum transfer in lepton/hadron--hadron and lepton/hadron--nucleus interactions and high densities and temperature in heavy ion collisions. Hence, the most important problem of nuclear physics concerns the role of quarks in forming nuclear structure: how are nucleons bound inside nuclei and do quarks manifest themselves explicitly in the ground-state nuclei?

We argue that nucleons within nuclei are bound with each other via quark--quark interactions which lead to strong nucleon--nucleon correlation in contrast to the independent particle approach. With this aim we propose so-called Strongly Correlated Quark Model (SCQM) \cite{Mus1, Mus2, Mus3} of hadron structure which is briefly described in Section 2 then in Section 3 applied to build the nuclear structure. It turned out (Section 4) that the received  geometry of nuclear structure corresponds to symmetry of the face-centered cubic (FCC) lattice. The SCQM together with the FCC lattice model developed by N.D. Cook \cite{Cook1, Cook2, Cook3} are applied to describe the nuclear properties.
\section{Strongly correlated quark model}
According to QCD nucleons are composed of three valence quarks, gluon field and sea of quark-antiquark pairs. Quarks possess various quantum numbers: flavour (u, d), electric charge (+2/3, -1/3), spin (1/2) and color (Red, Green, Blue). Exchange particles mediating interactions between quarks are ``gluons'' possessing spin 1 and different colors:
\begin{equation}
 R\bar{G}, G\bar{R}, R\bar{C}, C\bar{R}, G\bar{C}, C\bar{G}, R\bar{R}, G\bar{G}, B\bar{B}
\end{equation}
From latter three gluons one can make two combinations
\begin{equation}
  \sqrt{\frac{1}{2}} (R\bar{R}-G\bar{G}), {\sqrt{\frac{1}{6}} (R\bar{R}+G\bar{G}-2B\bar{B})}.
\end{equation}
As a result there are eight linearly-independent combinations of gluons which can be made in various ways. However, derivation of nucleon properties from the first principles of QCD is still not the solved task. Hence, there is a variety of phenomenological models which can be united in two groups: current quark models and constituent quark models. In current quark models relativistic quarks having masses 5--10 MeV move freely inside the restricted volume or a bag. In constituent quark models the quark-antiquark in mesons and three quarks in baryons are non-relativistic, surrounded by quark-antiquark sea and gluon field.

Our approach is based on similarity of quarks to solitons \cite{Frenk, Raja}. Description of nucleons in the framework of SCQM in details is given in papers \cite{Mus1}. We start with the quark--antiquark pair which oscillate around their centre of mass because of destructive interference of their color fields.  For such
interacting $q\overline{q}$ pair located from each other on
distance $2x$, the total Hamiltonian reads
\begin{equation}
H=\left[ \frac{m_{q}}{(1-\beta ^{2})^{1/2}}+U(x)\right] +\left[
\frac{m_{\overline{q}}}{(1-\beta ^{2})^{1/2}}+U(x)\right] =H_{\overline{q}}+H_{q}.
\label{hamil}
\end{equation}
Here $m_{q}$ and $m_{\overline{q}}$ are the current masses of the valence quark and
antiquark, $\beta =\beta (x)$ is their velocity depending on
displacement $x$, and $U(x)=\frac{1}{2}V_{q\overline{q}}(2x)$, where  $V_{q\overline{q}}$ is the $q\overline{q}$ potential energy with separation $2x.$  Assuming that
\begin{equation}
2U(x)=\int_{-\infty }^{\infty }dz^{\prime }\int_{-\infty }^{\infty
}dy^{\prime }\int_{-\infty }^{\infty }dx^{\prime }\rho (x,{\mathbf{r}%
^{\prime }})\approx 2M_{Q}(x)
\label{pot}
\end{equation}%
where $M_{Q}(x)$ is the dynamical mass of the constituent quark.

W. Troost \cite{troost} demonstrated that the Hamiltonian (\ref{hamil}) corresponds to the breather (soliton--antisoliton) solution of Sine-Gornon equation.  He derived the effective potential $U(x)$ for this solution
\begin{equation}
U(x)=M\tanh ^{2}(\alpha x),
\label{poten}
\end{equation}
where $M$ is a mass of soliton/antisoliton and $\alpha$ is an adjusting parameter. Thus, we identify our potential of quark--antiquark interaction in hamiltonian (\ref{hamil}) with the potential of soliton--antisoliton interaction.

\begin{wrapfigure}{r}{0.32\textwidth)}
\vspace{-35pt}
\begin{center}
\includegraphics[width=0.31\textwidth]{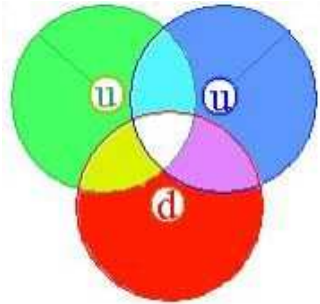}
\end{center}
\vspace{-30pt}
\caption{Nucleon made of three color quarks, R, G, B, surrounded by corresponding color fields.}
\vspace{-10pt}
\label{nucleon}
\end{wrapfigure}

\begin{figure}[ht]
\vspace{-10pt}
\begin{center}
\includegraphics[width=0.9\textwidth]{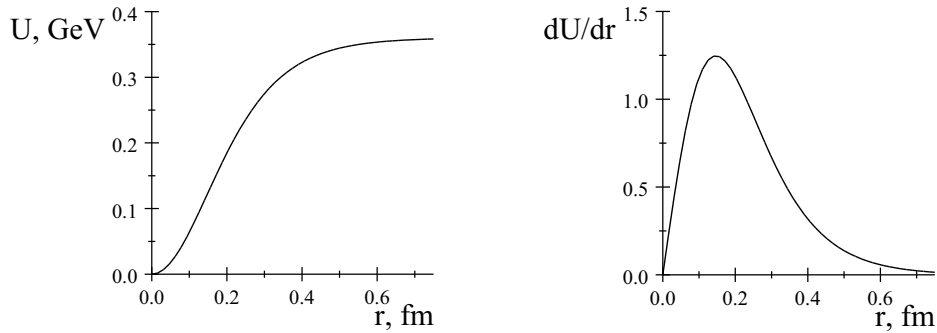}
\end{center}
\vspace{-15pt}
\caption{Left: Quark potential (Eq. \ref{pot}). Right: Interquark force. }
\vspace{-10pt}
\label{pot-force}
\end{figure}

Since quarks are members of the fundamental color triplet,
generalization to the 3-quark system (baryons, composed of Red,
Green and Blue quarks) is performed according to
$SU(3)_{color}$ symmetry: a pair of quarks has coupled
representations
\begin{equation}
3\otimes 3=6\oplus \overline{3}
\end{equation}
and for quarks within the same baryon only the $\overline{3}$ (antisymmetric)
representation is realized. Hence, an antiquark can be replaced by
two correspondingly colored quarks to create a color singlet baryon;
now destructive interference takes place between color fields of three
valence quarks (VQs). Putting aside the mass and charge differences
of valence quarks one may say that inside the baryon three quarks
oscillate along the bisectors of equilateral triangle under the potential (\ref{poten}). The larger separation of the color quarks is controlled by the linear growing confining potential.
All three quarks inside a nucleon are strongly correlated that is caused by the interaction (overlap) of their color fields (Fig. \ref{nucleon}). For convenience of perception the color fields are depicted as flat color circles. In QCD these interactions are mediated by gluons. For example, interaction between $R$ and $G$ quarks can be performed by the combination of gluons $\sqrt{\frac{1}{2}}(R\bar{G}+G\bar{R})$. Hereinafter we consider VQs oscillating on the $XY$ plane and assume that their spins point perpendicular to the plane of oscillations.
The parameters of the model, namely, the maximum displacement,
$x_{\max },$ and the parameters of the gaussian function, $\sigma
_{x,y,z},$ for hadronic matter distribution around VQ are chosen to
be $x_{\max }=0.64\ ${fm}$,\ \sigma _{x,y}=0.24\ ${fm}$,\ \sigma
_{z}=0.12\ ${fm}$.$ They are adjusted by comparison between the calculated
and experimental values of the total and
differential cross sections for $pp$ and $\overline{p}p$ collisions \cite%
{Mus3}. The mass of the constituent quark corresponding the value of the potential at maximum displacement is taken as
\begin{equation}
M_{Q(\overline{Q})}(x_{\max })=\frac{1}{3}\left( \frac{m_{\Delta }+m_{N}}{2}%
\right) \approx 360\ MeV,
\end{equation}
where $m_{\Delta }$ and $m_{N}$ are
masses of the delta isobar and nucleon correspondingly. The shape of the potential (\ref{poten}) and iterquark force are shown on Fig. \ref{pot-force}. As seen from the right plot, quark--quark coupling tends to zero at the origin of oscillation ("asymptotic freedom"), increases becoming maximal at intermediate values of displacement, and goes to zero at distances beyond the maximal displacement.
Although our description is classical it is justified by E. Schrodinger's approach in paper \cite{schrod} where he, analyzing motion of the gaussian wave packet for time dependent
Schrodinger equation for harmonic oscillator, demonstrated that this wave
packet moves in exactly the same way as corresponding classical oscillator.
 In our model VQ with its surroundings can be treated as
(nonlinear) wave packet and this wave packet really possesses soliton-like features.
 Because of plane oscillations of VQs and the flattened shape the hadronic
matter distribution around them, the 3-quark system, representing
baryons, is a non-spherical, oblate object. This
feature of nucleons plays an important role in the structure of
nuclei.

\section{Multinucleon Systems, Nuclei}

As shown in the previous section interaction between quarks within the nucleon arises owing to overlapping of their colour fields. The same overlapping mechanism of quark--quark interactions is responsible for nucleon--nucleon binding in nuclei. In this case different color fields of quarks belonging to the neighbour nucleons being overlapped create additional minima of the potentials at the maximal quark displacements in each nucleon with a small ($\sim$ 2 -- 8 MeV)  well depth. With regards to  the spin and flavor alignment of adjacent quarks, we should take
into account the fact that the multiquark states of 6, 9, 9, and 12
quarks in deuteron, $^{3}$H, $^{3}$He and $^{4}$He belong to the
completely antisymmetric representation of the $SU(12)$ group
which contains the direct product
\begin{equation}
SU(2)_{flavor}\otimes SU(2)_{spin}\otimes SU(3)_{color}.
\end{equation}
 That is, up to 12 quarks can
occupy the $s$-state. Some quark configurations in the above
multiquark systems built according to the group representations
correspond to, so-called, \textquotedblleft hidden
color\textquotedblright\ states as these can not be represented in
term of the free (color-singlet) nucleons. We restrict the
multiquark configurations only by the color-singlet
clusters--nucleons composing nuclei. In that way, nucleons will be bound
if the following rules are imposed on the linkage of two quarks:
\begin{quote}
 1) $SU(3)_{color}-$antisymmetric,\\
 2) $SU(2)_{isospin}-$antisymmetric,\\
 3) $SU(2)_{spin}-$symmetric.
\end{quote}

 Applying these rules one can construct any nucleus.  The three-nucleon system is formed by the linkage
of two quarks of each nucleon with quarks of two other nucleons
according to the above rules.
\begin{figure}[ht]
\vspace{-10pt}
\begin{center}
\includegraphics[width=0.8\textwidth]{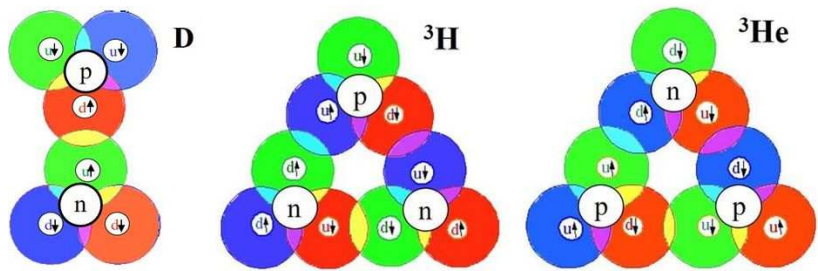}
\end{center}
\vspace{-15pt}
\caption{2-- and 3--nucleon systems. Quarks form in 3-nucleon systems the color quark loops. }
\vspace{-10pt}
\label{2-3-nucleon}
\end{figure}

Three-nucleon nuclei, namely $^{3}$H
and $^{3}$He, represent triangular configurations with quark loop and three quarks
at free ends (Fig. \ref{2-3-nucleon}). Completion of four-nucleon system,
$^{4}$He, from three-nucleon one, occurs by binding the free quark
ends in $^{3}$H ($^{3}$He) with the three quarks of an additional
proton (neutron) again in accordance with the above rules.
Here we should make a remark.
\begin{table}[!h]
\caption{Relation of binding energy per nucleon with quark loops and unbound quark ends }
{\begin{tabular}{@{}cccc@{}}
\hline
  {Nucleus} & {$E_{bind}$/nucl., MeV}  & {Quark loops}
  & {Unbound quark ends} \\
\hline
  $^{2}H$ & 1.11 & 0 & 4 \\
  $^{3}H$ & 2.83 & 1 & 3 \\
  $^{3}He$ & 2.57 & 1 & 3 \\
  $^{4}He$ & 7.07 & 4 & 0 \\ \hline
\end{tabular}
}
\label{tbl1}
\end{table}
 As seen from the Table \ref{tbl1}, the binding
energy per nucleon is minimal for the deuteron and maximal for the
$^{4}$He nucleus. This variability is due to the number of quark/color loops and the number of unbound quarks ends. Quark or color loops are created by the linkage of the quark ends of three nucleons, as in $^{3}$H and $^{3}$He. The more color loops the larger is the binding energy.
On the other hand the more unbound quark ends the less is the
binding energy. The maximal binding energy of $^{4}$He is due to the
presence of four color loops, binding all quark ends of the four
nucleons. The relationship between the binding energy and number of quark loops is closely related in turn to the additional potential well depth. The more quark loops the deeper the potential well. Exotic isotopes of $^{4}$He, $^{6}$He and $^{8}$He are
(loosely) bound systems due to the presence of color loops created
by di-neutrons bound with the protons of $^{4}$He core.  Removal of one of the neutrons composing a di-neutron destroys the color loop and the other neutron becomes unbound.

\begin{figure}[ht]
\vspace{-25pt}
\begin{center}
\includegraphics[width=0.8\textwidth]{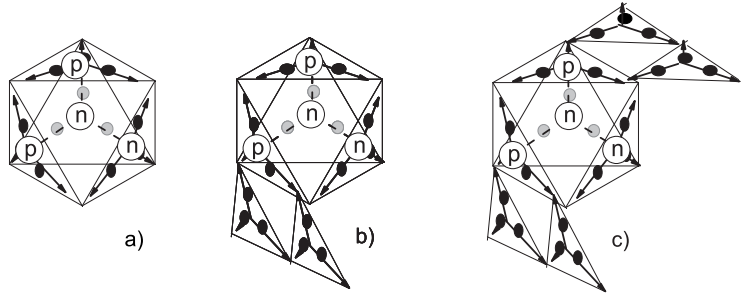}
\end{center}
\vspace{-25pt}
\caption{The core $^{4}$He nucleus (\textit{a}), and its exotic isotopes $^{6}
$He (\textit{b}) and $^{8}$He (\textit{c}); nucleons are depicted schematically as triangles.}
\vspace{-10pt}
\label{He-borr}
\end{figure}

\begin{figure}[ht]
\vspace{-25pt}
\begin{center}
\includegraphics[width=1.0\textwidth]{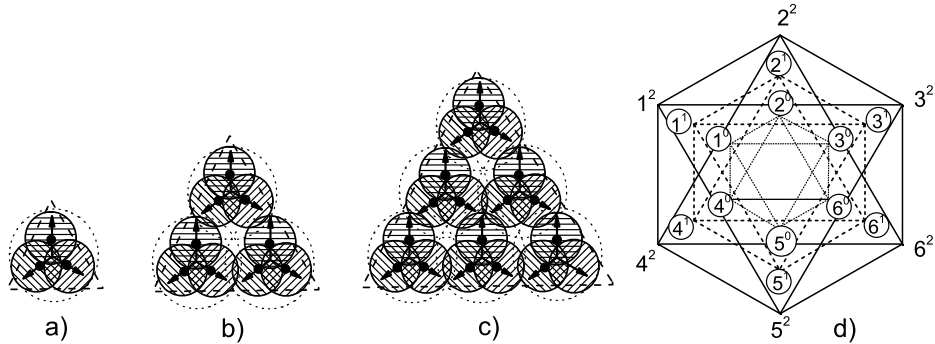}
\end{center}
\vspace{-25pt}
\caption{Building blocks of s- (a), p- (b) and d- (c) shells depicted as 3 nested octahedrons (d). Superscripts of vertex numbers indicate shells with $l = 0, 1, 2$.}
\vspace{-10pt}
\label{3-shells}
\end{figure}

Comparing geometrical shapes of three-nucleon systems and $^{4}$He one can conclude that the dimensions (rms-radii) of the formers should exceed the dimension of the latter, as three-nucleon systems are formed on a plane while four nucleons in $^{4}$He settle down on octahedron faces. Moreover, in the center of both three-nucleon systems and $^{4}$He there should be {\bf depression} or a {\bf hole} of the nuclear matter distribution. This peculiarity of 3- and 4-nucleon systems has been found out in model-independent analysis of electron scattering data performed by I. Sick and co-authors \cite{Sick87}. Both effects are a consequence of the non-spherical, oblate shape of the nucleons. Namely the oblate shape of nucleons leads to abrupt increase in dimension of halo nuclei. For example, in $^{6}$He and $^{8}$He two pairs of loosely bound neutrons linked to the protons stretch far from the octahedron--core $^{4}$He (Fig. \ref{He-borr}). For simplicity the nucleons are depicted as flat triangles.
Starting from the structure of the $^{4}$He all nuclei possess crystal-like structure (Fig. \ref{3-shells}). Indeed, in $^{4}$He a pairs of
(oblate) protons are located on the opposite faces of the
octahedron having a common vertex and a pair of neutrons --- on opposite faces of the another half of bipyramid. In this geometrical configuration
four nucleons are in $s$ state that corresponds to the first
\textit{s} shell of the shell model. Next, the $p$ shell can be
represented as a larger octahedron with two $^{3}$He triangles
instead of protons and two $^{3}$H triangles instead of neutrons.
The triangles are located parallel to empty
faces of the $^{4}$He octahedron, the free quark ends of these
triangles are coupled as in the $^{4}$He octahedron. This octahedron
with the nested $^{4}$He octahedron represents the nucleus of
$^{16}$O. The next shell with principal number $n=2$ is constructed
in the same manner, extending triangles beforehand by adding a row
of three protons to the row of two neutrons in $^{3}$H and a row of
three neutrons to the row of two protons in $^{3}$He (Fig. \ref{3-shells}). Again, these triangles are located in couples on opposite faces of an octahedron parallel to unoccupied faces of the
nested $p$ octahedron. Construction of the next
shells is performed in the same manner by extending triangles with
new rows of neutrons and protons. The nuclei built in accordance
with the model are found to exhibit symmetries that are isomorphic
with the independent particle description (shell model) of nucleon
states. The model reproduces not only $n$ shells but shell/subshell
structure implied by the wave equation of the shell model, at least
for $n\leq 2$ . For larger nuclei, however, the additional factor comes to play
the increasingly important role --- Coulomb repulsion of
protons. This is the reason why nuclei with Z $>$ 20 have excess
neutrons. At these values of Z the Coulomb repulsion force acting on
an additional proton at a specific position decreases the depth of
additional minimum of the quark potential.

Summing up, we emphasize that 3- and 4-nucleon configurations
play an important role in forming the bound multinucleon systems. According to our approach, quark loops are the basic blocks of binding of nucleons within nuclei, both stable and exotic (borromean). Namely, the formation of three-nucleon configurations is
responsible for {\bf the pairing--effect} of binding energies because
only two additional protons (neutrons) can form a quark loop with
one nuclear neutron (proton). Further, pairs of protons and pairs of
neutrons can form virtual alpha clusters within the nucleus. Hence, in our approach, in contrast to the independent particle (shell) model, {\bf protons and neutrons are strongly correlated}. Nevertheless, this strongly correlated system does not contradict quantum mechanics, inasmuch it is totaly anti-symmetrized according to the rules i -- iii in Sec. 3, and the equilibrium position of each nucleon obeys the uncertainty principle. Strong proton--neutron correlations modify nuclear magicity: not all magic numbers following from the shell model concern nuclei with (sub)shell closure. For example, though $^{10}$He should be (according to the shell model)  double--magic, such a bound state consisting of 2 protons and 8 neutrons has not been observed yet. According to our approach such a bound state does not exist and $^{8}$He is the last bound He isotope.
\section{From SCQM to FCC Lattice}
\subsection{FCC Lattice model}
Within nuclei constructed this way the nucleons aggregate into a
face-centered cubic (FCC) lattice with alternating spin and isospin
layers. It turns out that this arrangement is the basis of the
FCC-lattice model of the nuclear structure \cite{Cook1, Cook2,
Cook3}, developed more than 30 years ago. For finite nuclei the
FCC arrangement appears as a tetrahedron ($^{4}$He) and truncated
tetrahedrons (for larger nuclei). According to the FCC a nucleon's
principal number, \textit{n}, is a function of the nucleon's
distance from the center of the lattice --- leading to approximately
spherical shells for each consecutive \textit{n} eigenvalue:
\begin{equation}
n=(|x|+|y|+|z|-3)/2,
\end{equation}%
where $x,y,z$ are odd integers. The first shell (\textit{s} shell)
contains four nucleons with coordinates 111, -1-11, 1-1-1, -11-1.
The second shell (\textit{p} shell): 12 nucleons 31-1, 3-11, -311,
-3-1-1, 1-31, -131, 13-1, -1-3-1, -113, 11-3, 1-13, -1-1-3 and so
on\ldots\ The total angular
momentum value of a nucleon in the lattice%
\[
j=(|x|+|y|-1)/2
\]

is defined in terms of the distance of the nucleon from the spin axis of the
system -- leading to roughly cylindrical \textit{j} subshels within each
\textit{n} shell. The azimuthal quantum number%
\[
m=|x|/2
\]
is a function of the nucleon's distance from a central plane through
the lattice. We have thus demonstrated that
the FCC structure brings together shell, liquid-drop and cluster
characteristics, as found in the conventional models, within a
single theoretical framework. Unique among the various lattice
models, the FCC reproduces the entire sequence of allowed nucleon
states as found in the shell model. Correspondence between the FCC and shell model is not surprising because the geometrical shells of the lattice unambiguously reproduce the basic energy shells ($n$) that are a direct implication of the Schr\"{o}dinger equation.
\subsection{Combined SCQM--FCC: Nuclear properties}
According to SCQM--FCC an infinite nucleus (excluding Coulomb interaction) can be represented by {\bf alternating spin and isospin layers}. In FCC lattice the maximal number of the nearest neighbors to any nucleon is equal 12. This number, say, defines the maximally possible number of neutrons (isotopes) for any element. At the same time the quark--quark correlations resulting in the basic 3- and 4-nucleon configurations restrict nucleon--nucleon bonds prescribed by FCC lattice: among all lattice bonds only those are realized which form virtual 3-nucleon ( $^{3}H-$ and $^{3}He-$like) and 4-nucleon ($^{4}He-$like) configurations. Namely these configurations are responsible for the ``paring'' and ``even--even'' effects in nuclear binding energy. \\
{\bf Clustering:}
Though 3- and 4-nucleon configurations could be considered as clusters, we call them ``virtual'', as one or two nucleons can belong to adjacent clusters simultaneously. Examples of virtual clusters in nuclei can be represented by the following states: $^{6}Li \rightarrow {^{3}He} + {^{3}H}$;  $^{6}Li \rightarrow {^{4}He} + {^{3}H}$; $^{7}Li \rightarrow {^{4}He} + 2\times{^{3}H}$ and others. One can notice that the number of nucleons in clusters exceeds total number of nucleons in a nucleus, as some nucleons are considered more than once.  As to alpha--cluster model which has a long story, the real alpha--particles can not form (according to the SCQM) a stable nucleus, but it contains even number of protons and neutrons. Partly virtual alpha--clusters transform to the real ones in alpha--radioactive nucleus decay. Heavier virtual clusters, like $^{12}C$, can be a part of heavy nuclei. \\
{\bf Shell rearrangement:}
Formation of new shells rearranges inner shells. For example, in $^{12}C$ 3-quark planes of four nucleons of $s-$shell change their orientation in such a way, that they form with eight nucleons of $p-$shell  four virtual alpha--clusters. This rearrangement dilutes the shape of $s-$shell and reduces the highest nuclear density observed in $^{4}He$ to the normal, saturation nuclear density. \\
{\bf Separation energy:}
3-- and 4--nucleon configurations are responsible for larger values of nucleon separation energies in comparison with the average binding energy per nucleon, inasmuch as knocking out of one nucleon requires breaking of one quark loop and 4 quark loops in 3--nucleon and 4--nucleon configuration respectively.\\
{\bf Nuclear deformation:}
And all nuclei even those with shell closure are deformed. The shell closure nuclei $^{4}He$, $^{16}O$ and $^{40}Ca$ take the shape of octahedron. Neutron access leads to further deformation of nuclei. The mostly deformed are the nuclei with a large neutron to proton ratio.  Hence, light ``halo'' nuclei and all heavy and super-heavy nuclei are highly deformed. \\
{\bf Nuclear collective motion:}
 As nuclei possess the crystal--like structure, various collective motion, such as rotations, vibrations, shape oscillations are inherent in them. For example, the giant dipole resonance and scissor vibrations are a consequence of alternating proton--neutron layers.
\section{Conclusion}
The combined SCQM--FCC model gives an explanation of nuclear properties and experimentally observed phenomena. It composes most features of the conventional models and unifies them. And what is important --- it possesses the predictive power taking into account only symmetry considerations without binding energy estimation. Unquestionably, the model must include a quantitative consideration of the nuclear forces binding nucleons on the basis of quark--quark interactions that is a task of the further development of the model.  We hope that the proposed qualitative model is the initial step to the solution of the long standing problem of the nuclear structure.

\end{document}